\documentclass[12pt,pdflatex]{article}
\usepackage{amsmath,amssymb,amsfonts,color,graphicx,cite,color,soul}

\usepackage[usenames]{xcolor}
\usepackage[english]{babel}
\usepackage{cite}
\usepackage{dcolumn}

\newcommand{\sh}[1]{#1\hskip-7pt \diagup}

\usepackage{color}

\usepackage{graphicx}
\usepackage{epsfig}
\usepackage{amsmath,amsfonts,amssymb}
\pdfoutput=1
\usepackage[T1]{fontenc}
\usepackage[utf8]{inputenc} %

\newcommand{\KL}{\left(}
\newcommand{\KR}{\right)}
\newcommand{\KKL}{\left[}
\newcommand{\KKR}{\right]}
\newcommand{\KKKL}{\left\{}
\newcommand{\KKKR}{\right\}}
\newcommand{\svph}{\sin \frac{\varphi}{2}}
\newcommand{\cvph}{\cos \frac{\varphi}{2}}

\begin{document}
 \title{Muon $g-2$ through a flavor structure on soft SUSY terms}

\author{F.V. Flores-Baez$^{1}$\footnote{fflores@fcfm.uanl.mx} \and M. G\'omez Bock$^{2}$\footnote{melina.gomez@udlap.mx} \and M. Mondrag\'on$^{3}$\footnote{myriam@fisica.unam.mx}}


\maketitle 
\noindent
{\it \small{$^1$ FCFM, Universidad Aut\'onoma de Nuevo Le\'on, UANL.
Ciudad Universitaria, San Nicol\'as de los Garza, Nuevo Le\'on, 66450, Mexico\\
$^2$ DAFM, Universidad de las Am\'ericas Puebla, UDLAP.
Ex-Hacienda Sta. Catarina M\'artir, Cholula, Puebla, Mexico.\\
$^3$ Instituto de F\'isica, Universidad Nacional Aut\'onoma de M\' exico    
 Apdo. Postal 20-364, M\'exico 01000 D.F., M\'exico.}}\\
 
\begin{abstract}

\noindent
In this work we analyze the possibility to explain the muon anomalous
magnetic moment discrepancy within theory and experiment through
lepton flavor violation processes.  We propose a flavor extended MSSM
by considering a hierarchical family structure for the trilinear
scalar Soft-Supersymmetric terms of the Lagrangian, present at the
SUSY breaking scale.  We obtain analytical results for the rotation
mass matrix, with the consequence of having non-universal slepton
masses and the possibility of leptonic flavour mixing. The one-loop
supersymmetric contributions to the leptonic flavour violating process
$\tau \to \mu\gamma$ are calculated in the physical basis, instead of
using the well known Mass Insertion Method.  The flavor violating
processes $BR(l_i \to l_j \gamma)$ are also obtained, in particular
$\tau\to\mu\gamma$ is well within the experimental bounds. We present
the regions in parameter space where the muon g-2 problem is either
entirely solved or partially reduced through the contribution of these
flavor violating processes.

\end{abstract}

\section{Introduction}
\label{intro}


It is well known that in contrast to electric charge conservation,
lepton number conservation is not associated with a gauge symmetry. In
the Standard Model (SM), the spontaneous breaking of the electroweak
symmetry produces eigenstates of the remaining gauge group that are
not in general eigenstates of the mass matrix
\cite{Weinberg:1967tq,Weinberg:1972tu,Glashow:1961tr,Glashow:1970gm}. But
after diagonalization of the mass matrix, the electroweak coupling matrix is also
diagonal in the mass basis, therefore there is no possibility for
lepton flavor violation.  Certainly this is now in contradiction with
the experimental evidence on neutrino mixing \cite{Cleveland:1998nv,Fukuda:1998mi,Ahmad:2002jz,Ahn:2002up} and
also the possible LFV Higgs decay \cite{Khachatryan:2015kon} 
which forces the structure of the models beyond the SM.

The original structure of the SM with massless, and thus degenerate
neutrinos, implied separately $\tau, \mu, e$ number conservation.  In
particular, the processes $\tau^{\pm}\to l^{\pm} \gamma, (l=\mu^{\pm},
e^{\pm})$ through gauge bosons loops are predicted to give\footnote{A
  maximal mixing and a value of $\Delta^{2}_{32}\approx 3\times
  10^{-3}(eV/c^2)^2$ gives $\mathcal{B}(\tau\to\mu\gamma)\approx
  \mathcal{O}(10^{-54})$.}  very low rates \cite{Lee:1977tib}, even considering
the experimental evidence on neutrino oscillations \cite{Cleveland:1998nv,Fukuda:1998mi,Ahmad:2002jz,Ahn:2002up}. Under
this evidence the amplitudes for the Lepton Flavor Violation (LFV)
processes at low energy are suppressed by an inverse power of the large
Majorana mass scale $M_{I}$ used in the well-known seesaw model
\cite{Yanagida:1979as,GellMann:1980vs}, which explains naturally the small masses for the
active left-handed neutrinos.  On the other hand, the experimental
bounds for the branching ratio $\mathcal{BR}(\tau^{\pm}\to \mu^{\pm}\gamma)$ \cite{Benitez:2010gm} set strong restrictions on
models of physics beyond the SM. 

A realistic possibility of physics beyond the SM is offered by
supersymmetry (SUSY), whose simplest realization containing the SM is
the Minimal Supersymmetric Standard Model (MSSM) (see for instance
\cite{Haber:2000jh}).  In terms of supersymmetry, the SM is embedded
in a larger symmetry which relates the fermionic with the bosonic
degrees of freedom.  As a consequence of this higher symmetry, the
SUSY framework stabilizes the electroweak scale, provides us with dark
matter candidates, as well as with a greater possibility of unification of all
fundamental interactions and a solution to the hierarchy problem.

The discovery of the Higgs boson
\cite{Aad:2012tfa,ATLAS:2013mma,Chatrchyan:2012ufa,Chatrchyan:2013lba}
and the search for sparticles at the LHC, have modified the parameter
space of supersymmetry as a near electroweak (EW) scale model \cite{CMS:2014mia,
  Buchmueller:2012hv,Heinemeyer:2011aa,Buchmueller:2013rsa}.  The
MSSM, as the first minimal supersymmetric extension of the SM, was
conceived to be near to the electroweak  scale, in order to set
viable phenomenological scenarios to analyze with available
ex\-pe\-ri\-men\-tal data. 
One important issue to be considered was the experimental absence of
Flavor Changing Neutral Currents (FCNC), which lead to the simplifying
assumption of universality in the sfermion scalar masses, keeping the
desired good behavior of FCNC's ({\it i.e.} bounded) and in addition,
reducing the number of free parameters.

The Constrained Minimal Supersymmetric Standard Model (CMSSM) was conceived
under the assumption of Grand Unified Theories (GUT) structures. It
considers in particular universal sfermion masses and alignment of the
trilinear soft scalar terms, $A_{f,ij}$ to Yukawa couplings at the
unification scale \cite{Martin:1997ns,AguilarSaavedra:2005pw}.
Nevertheless, neutrino oscillations made it imperative to reconsider the
flavour structure in the theoretical models.

The most recent LHC data points to a heavy spectrum for some of the
SUSY particles in the case this constrained model were realized in
nature. The relation between the Higgs mass and the fermions and sfermions masses
in supersymmetric models indicate either higher stops masses or large
mixture within stops \cite{Hahn:2013ria}.
It is the squark sector, and particularly the stop and gluino, which
tend to lift the mass scale of the MSSM \cite{Aad:2015iea,
  Buchmueller:2012hv,Heinemeyer:2011aa,Buchmueller:2013rsa,Kitano:2013snk}. However,
for the slepton sector the LHC data for the exclusion bounds are less
restrictive and masses may still be below the TeV scale
\cite{Aad:2014vma}.\footnote{See also:\\ {\tiny
    https://atlas.web.cern.ch/Atlas/GROUPS/PHYSICS/CombinedSummaryPlots/SUSY/ATLAS\_SUSY\_Summary/ATLAS\_SUSY\_Summary.png}}.
On the other hand, we could go beyond the constrained MSSM and explore
other possibilities for the flavor structure.  It is thus very
relevant to search for SUSY effects to indirect electroweak
precision processes through quantum corrections involving
superparticles, as the phenomenologically viable parameter space is
modified by experimental data, being this the main motivation of the
present work.

 In the MSSM the conventional mechanism to introduce LFV is through
  the Yukawa couplings of the right handed neutrinos, $N_{i}$, which
  generate off-diagonal entries in the mass matrices for sleptons
  through renormalization effects
  \cite{Borzumati:1986qx,Leontaris:1985pq}, particularly in the $LL$ block. Then the predicted rates
  for the $\tau\to\mu\gamma$ and $\mu\to e\gamma$ decays are not
  suppressed, and depend on the unknown Yukawa matrix e\-le\-ments,
  but they will not be detected in the future experiments if those
  elements are too small.
  
  In Ref. \cite{Arganda:2005ji} the authors 
  work also with these LFV processes, using the seesaw mechanism in the SM \cite{Arganda:2004bz} and supersymmetric models
  to extended neutrino and sneutrino sectors, and perform the one-loop calculation through the
  Renormalization Group Equations (RGEs) based on leading-log approximation. In the SM they use the neutrino-gauge loops, while in the supersymmetric model they 
  get the sneutrino-chargino loops.

  In Ref.~\cite{Hisano:1995nq} the authors
  noticed that there is another source of LFV, namely the left-right
  mixing terms in the slepton mass matrix, and that their
  contributions to the LFV processes can be large even when the off
  diagonal Yukawa couplings elements are small. Later, in a second
  paper\cite{Hisano:1995cp}, they incorporated the full mixing of the
  slepton masses and mixing in the neutralino and chargino sector and
  then performed a numerical diagonalization of the slepton mass
  matrices. An interesting result of their analysis is that the
  contribution from the left-right mixing is only important in the
  region where the mixing term is $m_{\tau}\mu\tan \beta \sim
  \mathcal{O}(\tilde{m}^2_{S})$ and they consider the trilinear soft terms $A_{E,ij}$
  contribution negligible. In the above expression $m_{\tau}$ is the tau mass,
  $\mu$ ($\mu_{susy}$ throughout this paper\footnote{In order to avoid confusion we denote the Higgsino mass free parameter as $\mu_{susy}$})
 is the Higgsino mass parameter; $\tan\beta=v_2/v_1$ is the
  ratio of Higgs vacuum expectation values (vevs) and $\tilde{m}^2_{S}$ is the supersymmetric scalar mass scale from the soft SUSY breaking.
  It is worth noting, though,  that this
  analysis was done with very different considerations on experimental
  data than those we have now. 
  
  A more recent work on this relation
  between the seesaw mechanism for neutrino mixing and charged lepton
  flavor violation is done in Ref.~\cite{Figueiredo:2013tea}, where a
  non-trivial neutrino Yukawa matrix at the GUT scale leads to
  non-vanishing LFV interactions at the seesaw scale through the RGEs. Another approach to the same
  pro\-blem has been done using high-scale supersymmetry in
  Ref.~\cite{Moroi:2013vya},
  where the Majorana mass matrix of right-handed neutrinos is taken to
  be diagonal and universal, while the neutrino Yukawa matrix is
  proportional to the neutrino Pontecorvo-Maki-Nakagawa-Sakata (PMNS)
  mixing matrix $U_{PMNS}$, and the product of the left and right
  handed neutrino masses is $y_{\nu, Ij}=\frac{\sqrt{2M_{N_R}
      m_{\nu_{L,I}}}[U_{PMNS}]_{Ij}}{v\sin \beta}$.
 
 This neutrino Yukawa matrix, which would be present in low energy
  phenomenology,
  changes also with the RGE running of the soft SUSY breaking parameters.
  This scheme of FV was proposed in
    Ref.~\cite{Moroi:1995yh}, where small off-diagonal
    elements of the slepton mass matrix are considered  and, in the interaction basis, the FV
processes are restricted by using these off-diagonal elements
    as free parameters; here the trilinear coupling is considered to be zero, $A=0$. 
    In Ref. \cite{Calibbi:2015kja} the trilinear coupling $A_0$ is considered only for the LR flavor mixing term, 
    in the LR term of the corresponding slepton $A_0$ is set it to zero. 
  There is also a more general phenomenological work considering
  non-diagonal LL, RR and EL blocks of sfermion mass matrices which are
  parameterized as a sfermion mass product and a free parameter for each matrix element in order to 
  do a numerical e\-va\-lua\-tion of the processes
  in the mass basis \cite{Arana-Catania:2013nha}, having all the elements of the $6 \times 6$ 
  sfermion mass matrix as parameters that might be constrained by the LFV processes. 
  Recent analysis of these general FV contributions are done in \cite{Arana-Catania:2014ooa,Arganda:2015uca}.
    This general sfermion mass matrix, 
although complete, implies a considerable increase in the number of
parameters. Nevertheless, the authors found in seven different possible
scenarios an upper bound for their off-diagonal
parameter.
We must say here that in most of the literature, although the calculation is done in a physical basis, 
what is done is a diagonalization of $2\times2$ blocks flavor sleptons and they still consider a flavour mixing parameter, 
which is off-diagonal on the mass matrix and is used as coupling in the MIA method, so their physical basis means that instead of using
the interaction basis states $\tilde{l}_{i,L}, \tilde{l}_{i,R}$, they use $\tilde{l}_{i,1}, \tilde{l}_{i,2}$ with $i=1,2,3$ as flavors.\\

There is as well work on supersymmetric models where
$R-parity$ violation is considered in the allowed superpotential
operators \cite{Dreiner:2012mx}, with the consequence 
of having LFV couplings directly present in the model.\\ 

A very important issue to be considered when lepton flavor mixing is
allowed is the extra contribution to the anomalous magnetic moment of
the muon. 
The experimental value of the $g-2$ is another
element of the electroweak (EW) precision data which has not been
completely explained by the SM \cite{Jegerlehner:2009ry,Miller:2012opa,Benayoun:2014tra}, 
despite the efforts that have been made for improving the hadronic contribution
calculations \cite{Bodenstein:2013flq,Goecke:2011bm, Davier:2010nc},
the dominant source of uncertainty in the theoretical prediction.  It
is well known that the main MSSM contribution to $g-2$ (we will call it $a_{\mu}$), involves
neutralino-slepton and chargino-sneutrino loops \cite{Moroi:1995yh, Martin:2001st}. 
Even the two-loop contribution in terms of $\tan \beta$ has been calculated in Ref.~\cite{Marchetti:2008hw}, where a reduction
was found of the discrepancy coming from an extra
contribution, within $14\%$ to $6\%$ of the one-loop MSSM
contribution, depending on different scenarios of parameter space.

In Ref. \cite{Badziak:2014kea} the supersymmetric calculation of $a_{\mu}$ has been updated
con\-si\-de\-ring both the chargino-sneutrino loop and the
neutralino-smuon loop. It was found that the chargino-sneutrino loop
dominates, especially in the case where all the scalar masses are
degenerate and, on the other hand, when the $\mu_{susy}$ parameter is large,
then $\tilde\chi^0-\tilde{\mu}$ could be enhanced. 
There has also been work done relating the parameters for g-2 anomaly, flavour violation, and $h\to \gamma\gamma$ in \cite{Giudice:2012pf}.  \\

In this work we present an analysis of a flavor violating extension of the MSSM (FV-MSSM) one-loop contribution to
$a_{\mu}$, which is driven by a LFV mechanism at tree level. The
LFV process $\tau \to \mu \gamma$ is used as an additional constraint
of the parameter space of the FV-MSSM. Our strategy for the implementation
of LFV consists in assuming that $A_{E,ij}$-terms follow a particular
structure in the context of textures. Furthermore, we take an ansatz
for the mass matrix for sleptons, allowing an exact diagonalization
\cite{GomezBock:2008hz} that results in a non-universal spectra for sfermion masses,
providing a clear way for having flavour mixing within sleptons at
tree level and the opportunity to work in the mass eigenstates
basis. Concerning the extra contribution to the anomaly coming from the FV-MSSM, we assume that
it comes mainly from the slepton-bino loop, $a_{\mu}^{\tilde{l}\tilde{B}}$, 
and we compare with the usual MSSM contribution from this loop.

The paper is organized as follows: In Sect. 2 we present the flavor
structure of sleptons from an ansatz for the trilinear scalar terms.
Then in Sect. 3 we show the one-loop analytical calculation of
$BR(\tau\to \mu \gamma)$.  In Sect. 4 we include the $a_{\mu}$
calculation and present the combined results in Sect. 5.  Finally, we discuss our conclusions in Sect. 6.

\section{Flavor structure in the soft SUSY breaking Lagrangian}
If supersymmetry exists in Nature it has to be broken, since there is
no evidence that these new particles exist at low energies
\cite{CMS:2014mia}. This symmetry breaking is achieved by the introduction of terms in
the Lagrangian, which break SUSY in such a way as to decouple the SUSY
partners from the SM particles, and at the same time stabilize the
Higgs boson mass to solve the hierarchy problem (see for instance
\cite{Martin:1997ns}).  The soft SUSY breaking Lagrangian in general
includes trilinear scalar couplings $A_{ij}^H$, as well as bilinear
couplings $B_{ij}$, scalar squared mass terms
$\tilde{M}^{2}_{\tilde{f}}$, and mass terms for the gauginos $M^2_i$.

Specifically, for the scalar fermion part of the {\it soft SUSY} terms
in absence of flavor mixing, as is considered in the MSSM, it will
have the following structure:
\begin{equation}
\mathcal{L}_{soft}^{f}=-\sum_{\tilde f_i} \tilde{M}_{\tilde{f}}^{2}\tilde{\bar{f}}_{i}\tilde{f}_{i}
-(A_{\tilde{f},i}\tilde{\bar{f_L}}^{i}H_{1}\tilde{f_R}^{i}+h.c),
\label{LsoftNFV}
\end{equation}
where $\tilde{f}$ are the scalar fields in the supermultiplet. In the
case of sfermions the $L,R$ are just labels which point out to the
fermionic SM partners, but as we are dealing with scalar fields they
have no longer left and right $SU(2)$ properties. In general they may
mix in two physical states by means of a $2 \times 2$ rotation matrix,
\[
\begin{pmatrix}\tilde{f}_{L} \\
\tilde{f}_R\end{pmatrix}
\leftrightarrow 
\begin{pmatrix}\tilde{f}_{1} \\
 \tilde{f}_2
\end{pmatrix}~.
\]
The first terms in (\ref{LsoftNFV}) contribute to the diagonal terms
of the $2\times 2$ sfermion mass matrix, while the second ones are
Higgs couplings with the different sfermions, and they contribute to
the off-diagonal $L-R$ terms of the mass matrix once the EW symmetry is
spontaneously broken.
As $i$ is a flavour index we can see that Eq. (\ref{LsoftNFV}) implies no flavor mixing.

In our case, where we do consider flavour mixing in the trilinear
  terms, $A_{f}^{ij}$ would be a general $3\times 3$ matrix, since we consider
  together the three flavours, with two scalar fields for each one. 
The complete fermionic trilinear terms are given as
\begin{equation}
\mathcal{L}^{soft}_{H\tilde{f}_{i}\tilde{f}_{j}}=
-A_{u}^{ij}\widetilde{\bar Q}_{i}H_{2}\widetilde{U}_{j}-
A_{d}^{ij}\widetilde{\bar Q}_{i}H_{1}\widetilde{D}_{j} -
A_{l}^{ij}\widetilde{\bar L}_{i}H_{1}\widetilde{E}_{j} + c.c.
\label{AFerm}
\end{equation}
Here $\widetilde{Q}_{i}$ is the squark doublet partner of the SM
$SU(2)$ left doublet and $\widetilde{U}_{j},\widetilde{D}_{j}$ are the
corresponding squarks singlets,  while $\widetilde{\bar L}_{i}$ is the
slepton doublet and $\widetilde{E}_{j}$ is the singlet.  In this work in particular,
we  only analyze the sleptonic part.
We will explain further in this paper the ansatz flavour structure we consider for this.
Once the EW symmetry breaking is considered, the above Lagrangian
(\ref{AFerm}) for the sleptonic sector takes the form
\begin{equation}
  \mathcal{L}_{H\tilde{f}_{i}\tilde{f}_{j}}=
  \frac{A_{l}^{ij}}{\sqrt{2}}\left[(\phi_{1}^{0}-i\chi_{1}^{0})\tilde{l}_{iR}^{*}\tilde{l}_{jL}-\sqrt{2}\phi_{1}^{-}\widetilde{l}_{iR}^{*}\widetilde{\nu}_{jL}
    +v_{1}\tilde{l}_{R}^{*}\tilde{l}_{L}\right] + h.c.  \nonumber 
\end{equation}
The soft terms are not the only contributions to the sfermion mass elements, the supersymmetric auxiliary fields $F$ and $D$ coming from the superpotential
also contribute to this mass matrix as we explain in the next section.

\subsection{Mass matrix for sfermions}
The contribution to the elements of the sfermion mass matrix come from
the interaction of the Higgs scalars with the sfermions, which appear
in different terms of the superpotential and soft-SUSY breaking terms
as is fully explained in \cite{Kuroda:1999ks,Okumura:2003hy}. In the
case of the slepton mass matrix, as we said before, the contributions
coming from {\it mass soft terms} are $\tilde{M}_{l,LL}^{2}$,
$\tilde{M}_{l,RR}^{2}$, from trilinear couplings after EW symmetry
breaking $A_{ij}^l$ and from the $F,D -$terms. We arrange them in
a block mass matrix as follows:
\begin{equation}
\tilde{M}_{l}^{2}=
\begin{pmatrix}
    \tilde{M}_{l,LL}^{2}+ F_{l,LL}+ D_{l,LL} &  A_{ij}^l+F_{l,LR} \\
    (A_{ij}^l+F_{l,LR})^\dag & \tilde{M}_{l,RR}^{2}+ F_{l,RR}+ D_{l,RR}~
\end{pmatrix}~.
\label{massdFD}
\end{equation}

The $F_{f}$ and $D_{f}$  are the auxiliary fields in the
supermultiplets, which are introduced to have the same bosonic and fermionic 
degrees of freedom, but are dynamical spurious \cite{Haber:2000jh}. 
The $F$-auxiliary field comes from the Higgs
chiral superfields and contributes to the mass matrix as follows:
\begin{eqnarray}
F_{l,LL,RR}&=& m_{l}^{2}(\tilde{\bar{l}}_L \tilde{l}_L+\tilde{\bar{l}}_R \tilde{l}_R)\notag \\
F_{l,LR} &= & m_{l}\mu_{susy} \tan\beta(\tilde{l}_{R}^{*}\tilde{l}_{L}+\tilde{l}_{L}^{*}\tilde{l}_{R})
 \label{FHtermass}
\end{eqnarray}

From the $D-$auxiliary fields which come from the scalar superfields of fermions we have the following mass terms:
\begin{eqnarray}
 D_{l,LL,RR}=-M_Z^2\cos 2\beta [(T_{3l}-s_W^2Q_l)\tilde{\bar{l}}_L\tilde{l}_L+s_W^2Q_l\tilde{\bar{l}}_R\tilde{l}_R]
\end{eqnarray}
where $l = e,\mu, \tau$.
The elements of the sleptons mass matrix Eq. (\ref{massdFD}), for the different flavors 
given by $i,j=e,\mu,\tau$ 
are 
\begin{eqnarray}
 m_{LL,l}^{2}& =&
\tilde{M}_{\tilde{L,l}}^{2}+m_{l_{L}}^{2}+\frac{1}{2}\cos2\beta(2M_{W}^{2}-M_{Z}^{2}),\\
m_{RR,l}^{2} & = &
\tilde{M}_{\tilde{E},l}^{2}+m_{l_{R}}^{2}-\cos2\beta\sin^{2}\theta_{W}M_{Z}^{2},\\
m_{LR,l}^{2} & = &
\frac{A_{l}v\cos\beta}{\sqrt{2}}-m_{l}\mu_{susy}\tan\beta.
\label{Aterm}
\end{eqnarray}

\subsection{Soft trilinear terms ansatz}

The lepton-flavor conservation is easily violated by taking
non-vanishing off-diagonal elements for each matrix, the size of such
elements is strongly constrained from the experiments. In the CMSSM, it is assumed that
the soft sfermion $2\times2$ mass matrices $\tilde{m}_{E}^{2},
\tilde{m}_{L}^{2}$ are proportional to the identity matrix, and $A_{e,ij}$
is proportional to the Yukawa matrix $y_{e,ij}$. With these soft terms
the lepton-flavor number is conserved exactly \cite{Hisano:1995nq}.
The non-universality of scalar masses has been studied in
supersymmetric mo\-dels in the context of string theory
\cite{Carvalho:2000xg}.  In Ref. \cite{King:2004tx}, the authors
assume a non-universality of scalar masses, through off-diagonal
trilinear couplings at higher energies.
In Refs.~\cite{Calibbi:2009ja,Vives:2010zza} a SU(3) flavor
symmetry is introduced, then by means of the Froggat-Nielsen mechanism
the associated flavon fields acquire vevs, which upon spontaneous symmetry
breaking  generate the couplings which mix flavours.

In the present work, we assume $\tilde{m}_{RR,l}^{2} \approx
\tilde{m}_{LL,l}^{2} =\tilde{m}_{S}^{2}$ but we
propose that there is a mixing of two of the scalar lepton families in the $LR$ mass terms.
This mixing may  
come from a discrete flavor symmetry, as could be the extension of the
SM with $S_3$ \cite{Kubo:2003iw,Mondragon:2007af,Canales:2013cga}, or
supersymmetric models with $Q_6$
\cite{Kubo:2012ty,Gomez-Izquierdo:2013uaa,Ishimori:2010au,Babu:2011mv}, which have the fermions
assigned to doublet and a third family in a singlet irreducible representations. 
In order to analyze the consequences of this flavor structure we construct an ansatz for the trilinear terms $A_t$. Our procedure is similar to
the work done in Ref. \cite{DiazCruz:2001gf} for FCNC's in the quark sector through an ansatz of soft-SUSY terms. 
In our case we consider the whole two families contributions 
and of the same order of magnitude, having the following form for the trilinear term:
\begin{equation}
 A_{l}=
\begin{pmatrix}
  0 & 0 & 0 \\
  0 & w & y \\
  0 & y & 1
\end{pmatrix}
A_{0} \label{BLO} .
\end{equation}

In this case one could have at tree level the selectrons in a singlet
irrep., decoupled from the other two families of sleptons.  This would
give rise to a $4\times 4$ matrix, diagonalizable through a unitary
matrix $Z_{\tilde{l}}$, such that
$Z_{\tilde l}^{\dag}\tilde{M}_{l}^{2}Z_{\tilde l}=\tilde{M}_{diag}^{2}$. 

Since we assumed that the mixing is in the smuons and staus only and
the selectrons are decoupled, the remaining $4 \times 4$ smuon-stau
mass matrix will have the following form:
\begin{equation}
\tilde{M}_{\mu -\tau}^{2}=\left(
\begin{array}{cccc}
  m_{LL,\mu}^{2} & X_{m} & 0  & A_{y} \\
   X_{m} & m_{RR,\mu}^{2} & A_{y} & 0 \\
  0 & A_{y} & m_{LL,\tau}^{2} & X_{t} \\
   A_{y} & 0 & X_{t} & m_{RR,\tau}^{2}
\end{array}\right),
\label{m2LObloq}
\end{equation}
where 
\begin{align}
A_{y}&=\frac{1}{\sqrt{2}}yA_{0}v\cos\beta, \notag\\
X_{m}&=\frac{1}{\sqrt{2}}wA_{0}v\cos\beta - \mu_{susy} m_{\mu}\tan\beta,\notag\\
X_{t}&=\frac{1}{\sqrt{2}}A_{0}v\cos\beta - \mu_{susy} m_{\tau}\tan\beta. 
\end{align}
This way we will have  physical non-degenerate slepton masses.\footnote{We assign the label $\tilde{\tau},\tilde{\mu}$ 
to the masses to show the relation to the non-FV sleptons.}

\begin{eqnarray}
m^{2}_{\tilde{\tau}_{1,2}}& = & \frac{1}{2}(2 \tilde{m}_S^2-X_{m}-X_{t}\pm R)\nonumber\\
m^{2}_{\tilde{\mu}_{1,2}}& = & \frac{1}{2}(2 \tilde{m}_S^2+X_{m}+X_{t}\pm R)
\end{eqnarray}
where $R=\sqrt{4 A_y^2+\left(X_{t}-X_m \right)^2}$

We may  write the transformation which diagonalizes the mass matrix as in Ref.\cite{GomezBock:2008hz}, as a $4\times4$ rotation matrix for sleptons 
$Z_{\tilde{l}}$, which is in turn  a $2\times2 $ block matrix
$ Z_{\tilde l}^{\dag}\tilde{M}_{\mu -\tau}^{2} Z_{\tilde l}=\tilde{M}_{l,diag}^{2}$, explicitly having
 the form
\begin{equation}\label{rotationZ}
Z_{\tilde{l}}=\frac{1}{\sqrt{2}}\left(
\begin{array}{cc}
\Phi & -\Phi\\
\Phi\sigma^3 & \Phi\sigma^3
\end{array}
\right) ~,
\end{equation}
\vspace{1cm}
where $\sigma_3$ is the Pauli matrix and
\begin{equation}
\Phi=\left(
\begin{array}{cc}
-\sin\frac{\varphi}{2} & -\cos\frac{\varphi}{2}\\
\cos\frac{\varphi}{2} & -\sin \frac{\varphi}{2}
\end{array}
\right).
\end{equation}
The non-physical states are transformed to the physical eigenstates by
\begin{equation}
\begin{pmatrix}
\tilde{\mu}_{L} \\
  \tilde{\tau}_{L}\\
  \tilde{\mu}_{R}\\
  \tilde{\tau}_{R}
\end{pmatrix}
=\frac{1}{\sqrt{2}}\left(
\begin{array}{cccc}
 -\sin \frac{\varphi}{2} & -\cos \frac{\varphi}{2} &  \sin \frac{\varphi}{2} & \cos \frac{\varphi}{2}\\
  \cos \frac{\varphi}{2} & -\sin \frac{\varphi}{2} &  -\cos \frac{\varphi}{2} & \sin \frac{\varphi}{2}\\
  -\sin \frac{\varphi}{2} & \cos \frac{\varphi}{2} &  -\sin \frac{\varphi}{2} & \cos \frac{\varphi}{2}\\
  \cos \frac{\varphi}{2} & \sin \frac{\varphi}{2}&  \cos \frac{\varphi}{2} & \sin \frac{\varphi}{2}\\
\end{array}
\right)
\begin{pmatrix}
  \tilde{l}_{1}\\
  \tilde{l}_{2}\\
  \tilde{l}_{3}\\
  \tilde{l}_{4}
\end{pmatrix}
= Z_{\tilde l}
\begin{pmatrix}
  \tilde{\mu}_{1}\\
  \tilde{\tau}_{1}\\
  \tilde{\mu}_{2}\\
  \tilde{\tau}_{2}
\end{pmatrix}
 \label{rotationB}
\end{equation}
\noindent where

\begin{eqnarray}
\tan\varphi & =& \frac{2A_y}{X_m -X_t},\nonumber \\
& & \nonumber \\
\end{eqnarray}

In the case of the MSSM without slepton mixing
we would need to revert the similarity transformation performed as $Z_{\tilde l}\tilde{M}_{l,diag}^{2}(y=0)Z_{\tilde l}^{\dag}=M_{\tilde{\mu},\tilde{\tau}}$, 
vanishing also the mixing parameter, $y=0$. 
Then we will get a diagonal by blocks matrix, where the two $2\times2$ bloques are the mass matrix for smuons and staus, respectively which can in turn be diagonalize
separately as in the usual MSSM, obtaining the two sleptons physical states $\tilde{l}_1,\tilde{l}_2$ for each flavor that we
identify with the MSSM slepton eigenstates. The masses for the smuons would then be the usual ones, 
\begin{equation}\label{NOFVmass}
 m_{\tilde{\mu}_{1,2}}^2 = \tilde{m}_S^2-\frac{1}{2}M_Z^2 \cos 2\beta\pm \sqrt{M_Z^2 \cos 2 \beta(-\frac{1}{2}+2s_w^2)^2+4X}~,
\end{equation}
where $X=A_0 \frac{v\cos\beta}{\sqrt{2}}-\mu_{susy}\tan\beta m_{\mu}$.

\subsection{Neutralino-lepton-slepton interaction}
We assume the usual MSSM form of neutralinos as a mixing of the fermionic part of vector superfields, {\it i.e.} gauginos and Higgsinos.
The symmetric mass matrix for neutralinos is given  by
\begin{displaymath}
M_N=\left(\begin{array}{c c c c}
M_1&0&-M_Z\sin{\theta_W}\cos{\beta}&M_Z\sin{\theta_W}\sin{\beta}\\
*&M_2&M_Z\cos{\theta_W}\cos{\beta}&-M_Z\cos{\theta_W}\sin{\beta}\\
*&*&0&-\mu_{susy}\\
*&*&*&0\end{array}\right).
\label{masaneutralinos}
\end{displaymath}
The diagonalization of the mass matrix implies transformation of the neutralinos as
\begin{displaymath}
\left( \begin{array}{c}
\tilde\chi^0_1\\
\tilde\chi^0_2\\
\tilde\chi^0_3\\
\tilde\chi^0_4 \end{array}\right) = 
\left(\begin{array}{cccc}
\eta_1&0&0&0\\
0&\eta_2&0&0\\
0&0&\eta_3&0\\
0&0&0&\eta_4\end{array}\right)(\Theta_N)\left(\begin{array}{c}
\tilde B^0\\
\tilde W^0\\
\tilde H_1^0\\
\tilde H_2^0\end{array}\right).
\end{displaymath}
In the rotation matrix
$\eta$ is a diagonal matrix, whose elements $\eta_j$ are introduced in such a way as  to change the phase of
those neutralinos whose eigenvalues become negative after 
diagonalization, {\it i.e.}
$\eta_j=1$ for $m_{\chi^0_j}>0$  and $\eta_j=i$ for $m_{\chi^0_j}<0$.

The general interaction Lagrangian
for neutralino-fermion-sfermion in the MSSM is given as follows
\cite{Kuroda:1999ks}
\begin{eqnarray}\label{lagrangiano3}
\mathcal L_{\tilde\chi^0\tilde f f}&=&-\frac{g}{\sqrt
2} \sum_n^4\bigg\{\sum_{X=1}^3\bar{\tilde\chi}_n^0\left[l_n^{NfL}P_L+r_n^{NfL}P_R\right]f\tilde
f_X^* \nonumber\\
&& + \sum_{X=4}^6\bar{\tilde\chi}_n^0\left[l_n^{NfR}P_L+r_n^{NfR}P_R\right]f\tilde
f_X^*\bigg\}+ h.c.
\end{eqnarray}
where the $(l_n)$ and ($r_n$) are the left and right
fermion-neutralino couplings, respectively. In this expression the
$P_{L,R}$ are the ordinary chiral operators, and
the labels for the corresponding scalar superpartners of fermions are $L$ for sfermions $X=1,2,3$ and $R$ for $X=4,5,6$
 in the interaction basis and $g$ is the $U(1)$ coupling constant. 

The neutralino-fermion-sfermion couplings in equation (\ref{lagrangiano3}) are given by
\begin{eqnarray}
l_n^{NeL}&=&-\eta^*_n\lbrack
(\Theta_N)_{n2}+\frac{s_W}{c_W}(\Theta_N)_{n1}\rbrack ~,\\
r_n^{NeL}&=&\eta_n\frac{m_e}{M_W\cos{\beta}}(\Theta_N)_{n3} ~,\\
l_n^{NeR}&=&\eta_n^*\frac{m_e}{M_W\cos{\beta}}(\Theta_N)_{n3}~, \\
r_n^{NeR}&=&2\eta_n\frac{s_W}{c_W}(\Theta_N)_{n1} ~,\label{acoplam}
\end{eqnarray}
where  $\eta\Theta_N$ is the rotation matrix which diagonalizes the neutralino mass matrix \cite{Haber:1984rc}.\\

Now, considering the sleptons mass eigenstates given in (\ref{rotationB}) we rewrite the neutralino-lepton-slepton interaction Lagrangian as
\begin{eqnarray}
 \mathcal L_{\tilde\chi^0\tilde l l}&=&-\frac{g}{\sqrt2} \sum_n^4 \bar{\tilde\chi}_n^0\bigg\{C_{n+}^{NeLR}
 \left[\sin\frac{\varphi}{2}(\tau\tilde{\tau}_2^* -\mu\tilde{\mu}_1^* )+\cos\frac{\varphi}{2}(\mu\tilde{\tau}_2^* +\tau\tilde{\mu}_1^* ) \right]+\notag \\
 &&-C_{n-}^{NeLR}
 \left[\sin\frac{\varphi}{2}(\tau\tilde{\tau}_1^* -\mu\tilde{\mu}_2^* )+\cos\frac{\varphi}{2}(\mu\tilde{\tau}_1^* +\tau\tilde{\mu}_2^* ) \right]
 \bigg\}~,
\end{eqnarray}
where $C_{n\pm}^{NeLR}=C_{n}^{NeL}\pm C_{n}^{NeR}$ and $C_{n}^{NeL(R)}=l_{n}^{NeL(R)}P_L+r_{n}^{NeL(R)}P_R$.\\

So, we can see here that we directly  introduce the FV into the interaction Lagrangian avoiding the need of a mass insertion in the propagators of the loops.

\section{$BR(\tau \to \mu + \gamma$)}
In general, the way lepton flavor violation is introduced in
  calculations in the supersymmetric loops is using the approximation
  method called Mass Insertion Approximation (MIA) \cite{Gabbiani:1988rb,Hagelin:1992tc,Gabbiani:1996hi,
    Arana-Catania:2013nha}, which uses a Taylor
  expansion in a mass parameter \cite{Raz:2002zx} giving qualitative good results \cite{Dedes:2015twa}. Then the calculation is
  done in a non-mass eigenstate basis expanding around the universal squark masses \cite{Hagelin:1992ws}.  
  This method assumes that off-diagonal elements are small, which generates a strong restriction on the allowed SUSY parameters. 
  On the other hand, working in the interaction basis the number of loops to be calculated is reduced 
 to one, giving a simple analytical expression for the free parameters involved.
  Concerning flavour violation via neutrino and sneutrino mixing,
  including a right-handed neutrino \cite{Hisano:1995nq},
  the MIA method is used to compute the one-loop  amplitude for this process.\\
  
  In this paper, rather than using the MIA method, we work in a
    physical basis by diagonalizing exactly the complete mass matrix obtaining mixed flavour sleptons, introducing only two free parameters, which we
    reduce to one by considering $w=1$, assuming the soft
    trilinear term ansatz proposed in the previous section, Eq.~(\ref{BLO}).

\begin{figure}
\centering
\includegraphics[width=.65\linewidth]{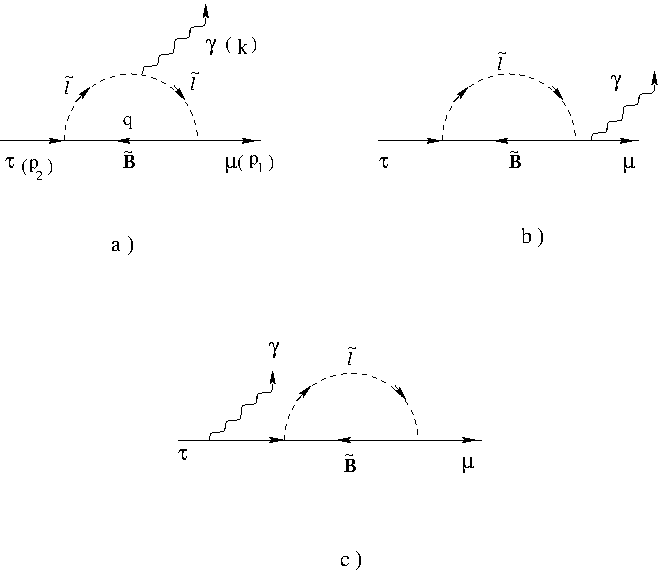}
\caption{One Loop diagrams in the LFV process  $\tau \to \mu \gamma$. The total amplitude is gauge invariant and finite in the UV region.}
\label{FigFVloop}
\end{figure}

We now use the couplings obtained to calculate FV processes to
establish the feasibility of the ansatz.  In particular, we calculate
the supersymmetric sfermion-neutralino one-loop contribution to the
leptonic flavor violation process $\tau \to \mu + \gamma$, which
corresponds to the Feynman diagram given in Fig.\ref{FigFVloop}.
The experimental bound to the branching ratio for this decay  at
$90\%$ C.L.~\cite{Benitez:2010gm} is
$\mathcal{BR}(\tau^{\pm}\to \mu^{\pm}\gamma) < 4.4\times
10^{-8}~.$
 
The loop diagrams shown in Fig.\ref{FigFVloop} are IR safe. A photon
is radiated either by a slepton inside the loop or by the external
lepton, all three diagrams are needed to achieve gauge invariance.
To simplify the expressions, we have assumed that the lightest
neutralino is mainly a Bino ($\tilde B$), although the procedure  can be
generalized to any type of neutralino. 

Considering the limit $M_1,M_2,\mu_{susy} \gg m_{Z}$ \cite{Haber:1984rc}, then
the lightest neutralino is mostly Bino $\tilde{\chi}_{1}^{0}\approx \tilde B$ then we take
$ (\Theta_N)_{1i}\approx\delta_{1i}  \label{BinoAprox}$ in Eq. (\ref{acoplam}). 
The mass eigenvalue for the lightest neutralino  is given by \cite{Martin:1997ns}
\begin{equation}
 m_{\tilde{N}_1}=M_1-\frac{m_Z^2 s_W^2(M_1+\mu \sin2\beta)}{\mu^2-M_1^2}+\cdots 
\end{equation}
Then this would be a Bino-like neutralino in the limit for numerical values $M_1<M_2 \ll|\mu_{susy}|$.
In this case the Bino-lepton-slepton coupling can be written as follows:
\begin{eqnarray*}
g_{\tilde{B}l_{i} \tilde{l}}= -\frac{g\tan\theta_{W}}{4}[S_{\tilde{B}l_{i}\tilde{l}}+P_{\tilde{B}l_{i}\tilde{l}}\gamma^{5}] \ ,
\end{eqnarray*}
where $\tilde{l}$ runs over the eigenstates $\tilde{l}_{1,2,3,4}$  given by Eq.(\ref{rotationB}).
For the $\tau \to \mu + \gamma $ decay the scalar and pseudoscalar couplings are given in Table \ref{Table:SP}.

\begin{table}
\begin{center}
\begin{tabular}{|c|c|c|c|c|}
\hline 
 $\tilde{l}$        & $\tilde{\mu}_{1}$ & $\tilde{\mu}_{2}$ & $\tilde{\tau}_{1}$ & $\tilde{\tau}_{2}$  \\ \hline
$S_{\tilde{B}\tau\tilde{l}}$& $3\svph$          & $\svph$           &  $\cvph$           & $3 \cvph$\\ \hline
$P_{\tilde{B}\tau\tilde{l}}$& $\svph$           & $3\svph$          & $3 \cvph$          &  $\cvph$\\ \hline
$S_{\tilde{B}\mu\tilde{l}}$ & $3 \cvph$         &  $\cvph$          & $-\svph$           & $-3\svph$\\ \hline
$P_{\tilde{B}\mu\tilde{l}}$ & $\cvph$           & $3 \cvph$         & $-3\svph$          & $ -\svph$\\ \hline
\end{tabular}
\caption{Scalar and pseudoscalar Bino-lepton-slepton couplings with lepton flavour mixing}
\label{Table:SP}
\end{center}
\end{table}

The total amplitude is  gauge invariant and free from UV divergences, as it should be, and it can be written in the 
conventional form,
\begin{align}
\mathcal{M}_{T}=\bar{u}(p_{1}) i\sigma^{\mu\nu}k_{\nu}\epsilon_{\mu}(E+ F\gamma^{5})u(p_{2})\ ,
\end{align}
where the one-loop functions $E$ and $F$ contain the sum of the contributions from
sleptons $\tilde{l}_{1,2,3,4}$ running inside the loop,
\begin{equation}\label{eq:EF}
E=5C\sum_{\tilde{l}} E_{\tilde{l}} \ , \qquad F=-3C\sum_{\tilde{l}} F_{\tilde{l}} \ , \qquad
C=\frac{ie g^{2}\tan^{2}\theta_{W}\sin\varphi}{2(16\pi)^2(m^2_{\tau}-m^2_{\mu})} .
\end{equation}

The functions $E_{\tilde{l}}, F_{\tilde{l}}$ are written in terms of
Passarino-Veltman functions and can be evaluated either by
LoopTools \cite{Hahn:1998yk} or by Mathematica using the analytical expressions for 
$C0$ and $B0$ \cite{Passarino:1978jh},
\begin{align}
  E_{\tilde{l}}&=\frac{\eta(\tilde{l})}{(m_{\tau}+m_{\mu})}
  \KKKL -\KKL 1+2m^{2}_{\tilde{l}}C0(m^{2}_{\tau},m^{2}_{\mu},0,m^{2}_{\tilde{l}},m^{2}_{\tilde{B}},m^{2}_{\tilde{l}})\KKR (m^2_{\tau}-m^2_{\mu}) \right.\nonumber\\ 
&\left.+ \frac{(m^2_{\tilde{l}}-m^2_{\tilde{B}})}{x}\left[ B0(m^{2}_{\mu},m^{2}_{\tilde{B}},m^{2}_{\tilde{l}})
    - B0(0,m^{2}_{\tilde{B}},m^{2}_{\tilde{l}}) - 
    x^2\KKL  B0(m^{2}_{\tau},m^{2}_{\tilde{B}},m^{2}_{\tilde{l}})-B0(0,m^{2}_{\tilde{B}},m^{2}_{\tilde{l}})\KKR \right] \right. \nonumber\\
&\left. + \KKL B0(m^{2}_{\tau},m^{2}_{\tilde{B}},m^{2}_{\tilde{l}})-B0(m^{2}_{\mu},m^{2}_{\tilde{B}},m^{2}_{\tilde{l}})
  \KKR \KL m_{\tau}m_{\mu}-2(m^2_{\tilde{l}}-m^2_{\tilde{B}})\KR \KKKR\nonumber\\
  &+(-1)^r \frac{8}{5}m_{\tilde{B}} \KKL B0(m^{2}_{\tau},m^{2}_{\tilde{B}},m^{2}_{\tilde{l}})-B0(m^{2}_{\mu},m^{2}_{\tilde{B}},m^{2}_{\tilde{l}})
  \KKR,
\label{eq:etaumu}
\end{align}
\begin{align}
  F_{\tilde{l}}&=\frac{\eta(\tilde{l})}{(m_{\tau}-m_{\mu})}
  \KKKL (m^2_{\tau}-m^2_{\mu})\KKL 1+2m^{2}_{\tilde{l}} C0(m^{2}_{\tau},m^{2}_{\mu},0,m^{2}_{\tilde{l}},m^{2}_{\tilde{B}},m^{2}_{\tilde{l}})\KKR \right. \nonumber\\
&\left.+  \frac{(m^2_{\tilde{l}}-m^2_{\tilde{B}})}{x}  \KKKL B0(m^{2}_{\mu},m^{2}_{\tilde{B}},m^{2}_{\tilde{l}})- B0(0,m^{2}_{\tilde{B}},m^{2}_{\tilde{l}}) - 
  x^2\KKL  B0(m^{2}_{\tau},m^{2}_{\tilde{B}},m^{2}_{\tilde{l}})-B0(0,m^{2}_{\tilde{B}},m^{2}_{\tilde{l}})\KKR \KKKR \right. \nonumber\\
&\left. + \KKL B0(m^{2}_{\tau},m^{2}_{\tilde{B}},m^{2}_{\tilde{l}})-B0(m^{2}_{\mu},m^{2}_{\tilde{B}},m^{2}_{\tilde{l}})
  \KKR  \KL m_{\tau}m_{\mu}+2(m^2_{\tilde{l}}-m^2_{\tilde{B}})\KR \KKKR,
\label{eq:eftaumu}
\end{align}
where we have defined the ratio $x=\frac{m_{\mu}}{m_{\tau}}$, and possible values of $r=1,2$ set by $\tilde{l}_r$, and the $\eta(\tilde{l})$ 
function as follows: $ \eta(\tilde{\tau}_{1,2})=-1$,  $ \eta(\tilde{\mu}_{1,2})=1$.

The differential decay width in the $\tau$ rest frame  reads
\begin{equation}
d\Gamma=\frac{1}{32\pi^{2}}[\frac{1}{2}\sum |\mathcal{M}|^{2}]\frac{|\vec{p}_{\mu}|d\Omega}{m_{\tau}^{2}}\ ,
\end{equation}
where $\vec{p}_{\mu}$ is the 3-vector of the muon. 
The branching ratio of the  $\tau\to \mu+\gamma$ decay  is given by the familiar expression,
\begin{eqnarray}
\mathcal{BR}(\tau\to \mu\gamma)=\frac{(1-x^2)^{3}m^3_{\tau}}{4\pi \Gamma_{\tau}}[|E|^{2}+|F|^{2}]\ .
\end{eqnarray}

\section{The MSSM and the muon anomalous magnetic moment $a_{\mu}$}

The anomalous magnetic moment of the muon $a_{\mu}\equiv
\frac{g-2}{2}$ is an important issue concerning electroweak precision 
tests of the SM.  The gyromagnetic ratio $g$, whose value $g=2$ is
predicted at lowest order by the Dirac equation, will deviate from this value 
when quantum loop effects are considered.  A significant difference
between the next to leading order contributions computed within the SM
and the experimental measurement would indicate the effects of new
physics. 

The experimental value for $a_{\mu}$ from the Brookhaven experiment
\cite{Bennett:2006fi} differs from the SM prediction by about three standard
deviations. In particular, in Ref.\cite{Jegerlehner:2009ry} it is found that the
discrepancy is
\begin{equation}\label{Deltag-2}
\Delta a_{\mu}\equiv a_{\mu}^{Exp}-a_{\mu}^{th}= (287\pm 80)\times 10^{-11},
\end{equation}
where $a_{\mu}^{th}$ is the theoretical anomalous magnetic moment of
the muon coming only from the SM.

Three generic possible sources of this discrepancy have been pointed
out \cite{Freitas:2014pua}. The first one is the measurement itself, although
there is already an effort for measuring $a_{\mu}$ to 0.14 ppp
precision \cite{Venanzoni:2012qa}, and an improvement over this measurement is planned at the J-Parc muon g-2/EDM experiment
\cite{Mibe:2011zz} whose aim is to reach a precision of 0.1 ppm.

The second possible source of discrepancy are the uncertainties in the
evaluation of the non-perturbative hadronic corrections that enter in
the SM prediction for $a_{\mu}$.  The hadronic contribution to
$a_{\mu}$ is separated in High Order (HO) and Leading Order (LO)
contributions. The hadronic LO is under control, this piece is the
dominant hadronic vacuum polarization contribution and can be
calculated with a combination of the experimental cross section data
involving $e^+e^-$ annihilation to hadrons and perturbative QCD
\cite{Davier:2010nc}.  The hadronic HO is made of a contribution at ${\cal O}
(\alpha^3)$ of diagrams containing vacuum polarization insertions
\cite{Hagiwara:2006jt,Krause:1996rf} and the very well known hadronic
Light by Light (LbL) contribution, which can only be determined from
theory, with many models attempting its evaluation \cite{Williams:2013tia,Nyffeler:2013lia}. 
The main source of the theoretical error for $a_{\mu}$ comes from LO and LbL
contributions. It is worth mentioning that the error in LO can be
reduced by improving the measurements, whereas the error in LbL
depends on the theoretical model.

The third possibility comes from loop corrections from new particles
beyond the SM. There have already many analyses been done in this direction
(see for instance
\cite{Hisano:1995nq,Vicente:2015cka,Nakamura:2015sya}).  

To calculate one-loop effects to g-2, for general contributions coming from different kind of particles 
Beyond the SM, there is a numerical code built using {\it Mathematica} \cite{Queiroz:2014zfa}.\\

The supersymmetry contribution to g-2, $a_{\mu}^{SUSY}$, was first computed by
  Moroi Ref.~\cite{Moroi:1995yh} and recently updated in
  Ref.~\cite{Endo:2013lva}.  In these works the large $\tan \beta$
  scenario was studied, showing the dominance of the chargino-sneutrino
  loop over the neutralino-smuon loop, provided the scalar masses are
  degenerate, otherwise the $\mu_{susy}$ parameter (Higgsino mass parameter) must be large allowing an
  enhancement of the muon-neutralino loop ($\chi^0-\tilde{\mu}$). 
  It was also shown that in the interaction basis the dominant
  contributions are proportional to $\mu_{susy} M_1 \tan\beta$, then the sign
  and the size of the contribution to $a_{\mu}^{SUSY}$ depends on the
  nature of this product.  Hence, the supersymmetric contributions to
  the anomaly are determined by how these elements are assumed (see for
  instance \cite{Moroi:1995yh,Endo:2013lva}).  The results in
  the literature are usually obtained using the MIA approximation,
  however, there are some schemes where the work is done in the
  physical basis (e.g. \cite{Arganda:2015uca}).
  The difference with the MIA method is not only the change in basis, 
  but the restriction that is imposed a priori that some elements in the mass matrix are considered small compared to the diagonal ones.\\

There has been research toward an MSSM explanation to the $g-2$
  discrepancy related to LFV as in \cite{Chacko:2001xd,Kersten:2014xaa},
  since there is a correspondence between the diagrams in the MSSM that
  contribute to the anomalous magnetic moment of the muon and the 
  diagrams that contribute to LFV processes. The process $\mu \to e+\gamma$ have been used to
  constrain lepton flavor violation and as a possible connection to $g-2$.

\begin{figure}
\centering
\includegraphics[width=.35\linewidth]{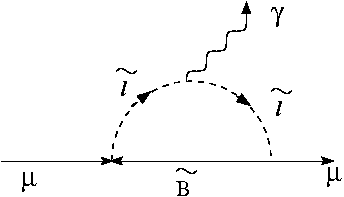}
\caption{Slepton contribution to $a_{\mu}$.}
\label{fig-slepton-cont}
\end{figure}

In this work we assume that there is room for an MSSM contribution to
$g-2$ through lepton flavor violation in the sleptonic sector. In particular, we search for the LFV process $\tau \to \mu+\gamma$
and calculate $g-2$ through a mixing of smuon and stau families, $a_{\mu}^{\tilde{l}\tilde{B}}$, Fig. \ref{fig-slepton-cont}. 
The ansatz proposed here avoids extra $\mu \to e+\gamma$ contributions.
To establish the restrictions on parameter space we consider a loose
constraint, $|a_{\mu}^{Exp}- a_{\mu}^{TH}|\leq 3.3\sigma$, where
$a_{\mu}^{TH}=a_{\mu}^{th}+a_{\mu}^{\tilde{l}\tilde{B}}$ 
indicates that the lepton flavor violation supersymmetric loop through
charged sleptons is not necessarily the only contribution to solve the discrepancy, Eq.(\ref{Deltag-2}).  
We also show the extreme case in parameter
space  where this loop contribution solves the discrepancy completely $|a_{\mu}^{Exp}- a_{\mu}^{TH}|\leq 1\sigma$.

When taking into account the slepton-bino flavor violation
  contribution to $g-2$, if the discrepancy is $\leq 1 \sigma$,
  it means that this contribution solves the whole $g-2$ problem.  In
  the opposite scenario,  $|a_{\mu}^{Exp}- a_{\mu}^{TH}|\approx
  3.3\sigma$  means that the slepton-bino loop gives no
  significant contribution to the discrepancy. In here we will look
  at a possible contribution to $g-2$ between both scenarios.

Using the LFV terms constructed previously we obtained the contribution to the anomalous magnetic
moment of the muon $a_{\mu}^{\tilde{l}\tilde{B}}$. Defining the
ratio $r= m_{\mu}/M_1$ and taking the leading terms when
$r^2\to 0$, and $M_1=m_{\tilde{B}}$ as the Bino mass.\\

In order to compute the SUSY contribution to the $g-2$ anomaly, we follow the method given in Ref.\cite{Peskin:1995ev}. 
All we have to do is to isolate the coefficient of the $(p_{1}+p_{2})^{\mu}$ term, in other words, computing the one-loop contribution, 
we can write the result  as follows:
\begin{align}
\bar{u}(p_{1})\Gamma^{\mu}u(p_{2})&=\bar{u}(p_{1})[  B(p_{1}+p_{2})^{\mu}+ _{\cdots}  ]u(p_{2})\nonumber\\
&=\bar{u}(p_{1})[\frac{\imath\sigma^{\mu\nu}}{2m_{\mu}}q_{\nu}]F_{2}(q^2)u(p_{2})+ _{\cdots}
\end{align}
where the ellipsis indicates terms that are not proportional to $(p_{1}+p_{2})^{\mu}$. Then the anomaly can be defined as
$\delta a_{\mu}=\frac{g-2}{2} = F_{2}(0)$ with $q=p_{2}-p_{1}$.\\
Keeping in mind that we require the magnetic interaction which is given by the terms in the loop process proportional to $(p_{1}+p_{2})^{\mu}$ we write it as
\begin{align}
&B(p_{1}+p_{2})^{\mu}\ ,\nonumber\\
\frac{g-2}{2}&=F_{2}(q^2\to 0)=-2m_{\mu}B.
\end{align}
Considering only these terms in the interaction and gathering them, the contribution of the flavour violation loop to the $g-2$ anomaly 
due to a given slepton $\tilde{l}$ reads
\begin{align}
\frac{g-2}{2}
&=\frac{g^2_{c}}{(4\pi)^2}(S^2_{\tilde{B}\mu,\tilde l}-P^2_{\tilde{B}\mu,\tilde l})\frac{2m_{\tilde{B}}m_{\mu}}{\Delta_{\tilde{l}\tilde{B}}}\left[ -\frac{1}{2}
-\frac{m^2_{\tilde{B}}}{\Delta_{\tilde{l}\tilde{B}}}
-\frac{m^2_{\tilde{B}}m^2_{\tilde{l}}}{\Delta_{\tilde{l}\tilde{B}}^2}\ln\left[\frac{m^2_{\tilde{B}}}{m^2_{\tilde{l}}}\right]\right]\nonumber\\
&+\frac{g^2_{c}}{(4\pi)^2}(S^2_{\tilde{B}\mu,\tilde l}+P^2_{\tilde{B}\mu,\tilde l})\frac{2m^2_{\mu}}{\Delta_{\tilde{l}\tilde{B}}}
\left[ \frac{1}{6}-\frac{m^2_{\tilde{B}}}{2\Delta_{\tilde{l}\tilde{B}}}
-\frac{m^4_{\tilde{B}}}{\Delta_{\tilde{l}\tilde{B}}^2}-\frac{m^4_{\tilde{B}}m^2_{\tilde{l}}}{\Delta_{\tilde{l}\tilde{B}}^3}
\ln\left[\frac{m^2_{\tilde{B}}}{m^2_{\tilde{l}}}\right]\right],
\end{align}
where $g^2_{c}=\frac{ \rm tan^2\theta_{w}g^2_{1}}{16}$, and $\Delta_{\tilde l \tilde B}=m^2_{\tilde l}-M^2_1$, having four
contributions with $\tilde{l}$ running from 1 to 4 with the values of
the couplings $S_{\tilde{B}\mu,\tilde l},P_{\tilde{B}\mu,\tilde l}$
are
given in Table \ref{Table:SP}.\\
This expression is equivalent to the one presented in
\cite{Stockinger:2006zn} and can be written using their notation as
can be found in Appendix B. 

The expression will be different from MIA because the off-diagonal
elements $LR$ are not explicit since we are in the physical basis.  In
the interaction basis, the $LR$ terms appear with explicit SUSY free
parameter dependence as they use directly the elements of the slepton
mass matrix. Exact analytical expressions for the leading one- and
two-loop contributions to g-2 in terms of interactions eigenstates can
be found in Refs.~\cite{Martin:2001st,Stockinger:2006zn}, and references
therein.  By taking these expressions in the limit of large
$\tan\beta$ and of the mass parameters in the smuon, chargino and
neutralino mass matrices equal to a common scale $M_{SUSY}$, the
results calculated in the mass-insertion approximation in the same
limit \cite{Moroi:1995yh} are reproduced from the complete forms given
in \cite{Stockinger:2006zn}.  We have explicitly checked that our
one-loop results when no LFV terms are present coincide with the
analytical expressions of ref.~\cite{Stockinger:2006zn}, and thus in
the appropriate limits also with the MIA expressions. Our expressions
for the contribution of the LFV terms to g-2 can be found in Appendix
B.\\

Here we take a flavour structure with no a priori restrictions on the size of the mass matrix elements other than two family mixing, 
and the restrictions come directly from the comparison with experimental data.

\begin{table}[hbt]
\renewcommand{\arraystretch}{1.5}
\begin{center} 
\begin{tabular}{|c|c|}
\hline
$\mu_{susy}\in[-15,15]$ \rm{TeV}& $A_0\in[50,5000]$ \rm{GeV}   \\
$\tilde{m}_S\in[50,5000]$ \rm{GeV} &$\frac{M_1}{\tilde{m}_S}\in[0.2,5]$ \rm{GeV} \\
 $\tan\beta\in[1,60]$&$w=-1$ , $y=1$\\
\hline
\end{tabular} 
\renewcommand{\arraystretch}{1.0}
\caption{The table shows the parameter space where the scan was performed. 
The values were taken at random for each variable within the bounds shown}.{\label{RunPar} }
\end{center}
\end{table}

\section{Results}
We now analyze the region in parameter space allowed by the
experimental bound on $BR(\tau \to \mu \gamma)$, taking into account
that the mixing parameters $w,y$ represent at most a phase, {\it i.e.}
the mixing terms in the $LR$ term of the mass matrix are of the same
order as $A_0$, see Eq.~(\ref{BLO}), in contrast with the MIA method where this terms are considered small compared with the diagonal
ones which is needed to apply the method.
In the parameter space region comprised by Table \ref{RunPar}, we are able to safely consider lepton
flavour mixing in trilinear soft terms of the MSSM, and constrained it at the current
experimental bounds $BR^{exp}(\tau \to \mu \gamma) < 4.4 \times 10^{-8}~$ \cite{Agashe:2014kda}.
Throughout parameter space we take $M_1 <|\mu_{susy}|$. 
We highlight the points where the $g-2$ is solved completely, shown in black in all figures. 
In order to ensure that the lightest neutralino is mostly Bino, 
we further assume for these points $M_1 \lesssim \frac{1}{3}|\mu_{susy}|$.

  We found for the parameter values given in Table \ref{RunPar} that
  the $BR(\tau \to \mu \gamma)$ is only partially restricted from experimental
  bound for  $\tilde{m}_S\lesssim 3200$ \rm {GeV}, also for $M_1 \lesssim 4.5$ {\it TeV}.
Table \ref{BoundBR} shows examples of  different sets of values for random parameters given within the range in 
Table \ref{RunPar}, consistent with the experimental bound on LFV
and that also solve entirely the g-2 discrepancy, in all these points the  Bino is considered as the LSP. 
From these sets of values it can be seen that the $g-2$ discrepancy can be solved within the FV-MSSM 
by different possible combinations of the parameters.\\

The difference between the experimental value and the SM prediction for the
anomalous magnetic moment, Eq.(\ref{Deltag-2}), 
gives $\sigma= 80 \times 10^{-11}$. As we have already explained  we distinguish between 
two possible ways the slepton contribution should be constrained,
depending whether the loop is dominant in FV-MSSM or not:
\begin{align}
  0< a_{\mu}^{\tilde{l}\tilde{B}}< 6.6 \sigma\,\,\,\, & 
\text{any contribution,} \label{6sigma}\\
  2.3\sigma < a_{\mu}^{\tilde{l}\tilde{B}}< 4.3 \sigma \,\,\,\, &
  \text{main contribution.} \label{4sigma}
\end{align}

It is important to mention that we take the points that solve for   ``any contribution'' as defined above (blue in graphs),
because we are aware that this is only one of the possible supersymmetric contributions to $g-2$.
 In a more general case we need to include the chargino-sneutrino contributions in order to have an entire picture of the parameter space.
 In the FV extension considered here this contribution will be the same as in the usual MSSM. 
 For a more complete treatment right-handed neutrinos should be considered, 
 together with LR mixing and the trilinear term.

\begin{table}[ht!]
\renewcommand{\arraystretch}{1.3}
\begin{center} 
\begin{tabular}{|ccccccc|} 
\hline
$BR(\tau\to\mu\gamma)$& $\alpha_{\mu}^{\tilde{l}\tilde{B}}$&$\tan\beta$  &$M_1 $\rm{(GeV)} &$\mu_{susy}$\rm{(GeV)} & $\tilde m_S$ \rm{(GeV)}& $A_0$(\rm{GeV}) \\ 
\hline
$3.06 \times 10^{-8}$& $2.17\times 10^{-9}$ &$15.4$& $1205 $&$7324.6 $& $457.6 $ & $145.2 $\\  \hline
$3.01 \times 10^{-8}$& $2.06\times 10^{-9}$ &$45$& $714 $&$10298 $& $991 $ & $1236.7$\\ \hline
$2.33 \times 10^{-8}$& $2.42\times 10^{-9}$ &$1.35$& $697 $&$-2832.2 $& $831.8 $ & $4003.5$\\ \hline
$2.22\times 10^{-8}$ &$3.13\times 10^{-9}$ &$30.7$ & $363.7 $ & $12554.7 $ & $832.2$ & $340 $ \\ \hline
$1.22\times 10^{-8}$ &$3\times 10^{-9}$ &$46.3$ & $509.7 $ & $4681.2 $ & $691 $ & $408.5$ \\ \hline
$2.06 \times 10^{-11}$& $2\times 10^{-9}$ &$45.6$& $2064 $&$9127 $& $1005.7 $ & $50.7 $\\  \hline
\end{tabular}   
\caption[]{Sample of parameter sets that solve entirely the muon $g-2$ discrepancy,
consistent with the experimental bound on 
  $BR(\tau\to\mu\gamma)$, calculated using random values of the parameters given in Table \ref{RunPar}. For all these sets the LSP is a Bino. }
\label{BoundBR}
\end{center}
\end{table}

 \begin{figure}[!htb]
\centering
\includegraphics[width=6cm]{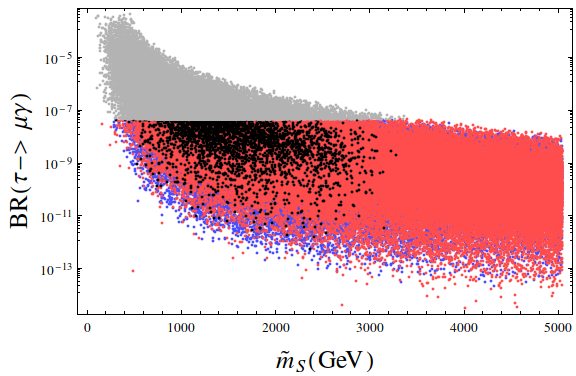}\includegraphics[width=6cm]{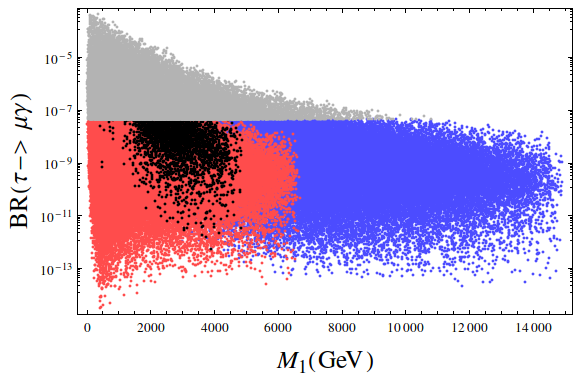}
\includegraphics[width=6cm]{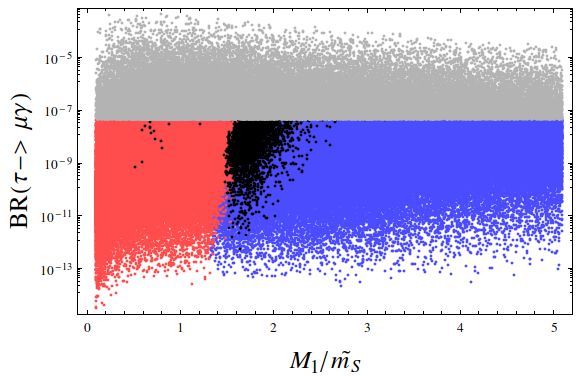}
\caption{The plots show the dependence on $ BR^{theo}(\tau \to \mu
  \gamma)$ on the SUSY scalar mass $\tilde{m}_S$  (left) and the Bino mass $M_{1}$ 
  (right) and on the ratio of them (down). The gray points are excluded by the experimental bound
  on $ BR(\tau \to \mu \gamma)$. The rest of the color code is shown explicit in Fig.\ref{grafs-amu-vs-M}, which separates
  ranges of FV contributions to $g-2$.}
\label{grafLogBR}
\end{figure}

\begin{figure}[!htb]
\centering
\includegraphics[width=6cm]{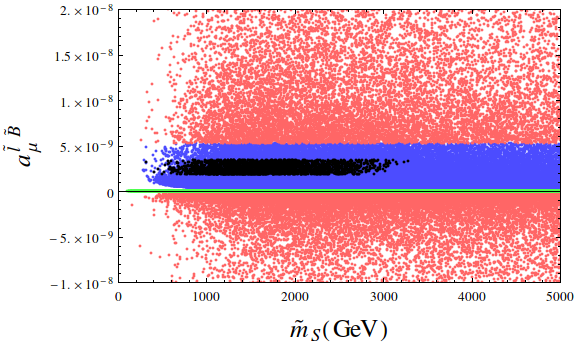}\includegraphics[width=6cm]{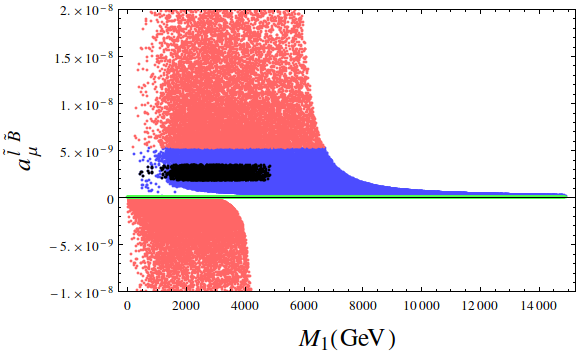}
\caption{The plots show the dependence of the value of our calculation for
    $a_{\mu}^{\tilde{l}\tilde{B}}$ with the SUSY scalar mass  (left)
    and the Bino mass (right). Here the color code used in Figs. \ref{grafLogBR}, \ref{graf-mu-vs-tanbmS} and \ref{grafg-2mSM1tB} is shown explicit as ranges of the $a_{\mu}^{\tilde{l}\tilde{B}}$.
    The green points correspond to no FV Bino-slepton loop, considering only the smuons in their mass
    eigenstates and $A_0=0$ the same as green points in previous figure (fig. \ref{graf-g2NFV-vs-mS}).}
\label{grafs-amu-vs-M}
\end{figure}

In Fig.~\ref{grafLogBR} we show the dependence of
the $BR(\tau \to \mu \gamma)$ on $\tilde m_S$ and on the Bino mass $M_1$, and it is shown the stringent restrictions for these masses. 
In Fig.~\ref{grafs-amu-vs-M} we show the value of
  $a_{\mu}^{\tilde{l}\tilde{B}}$ for different values of the Bino and the SUSY scalar mass, the color code is clear from this figure.
 The blue points correspond to the mass scale for which there is any
 contribution to the dis\-cre\-pan\-cy $a_\mu$ Eq.~(\ref{6sigma}). The black ones
  are those for which the discrepancy would be completely
 explained by the LFV contribution Eq.~(\ref{4sigma}), for these points we take $M_1 < \frac{1}{3}|\mu_{susy}|$ 
 (otherwise we just take $M_1 < |\mu_{susy}|$). The red points 
  are outside these ranges, {\it i.e.} are contributions non-compatible with experimental data
  of the muon g-2 anomaly to be solved.  The green points show the results obtained
  by taking $y=0$ in our ansatz, {\it i.e.} no FV, and calculating the smuon-Bino loops
  for $g-2$ with the smuons masses as given in Eq. (\ref{NOFVmass}) and considering a trilinear coupling as $A_0=0$.

\begin{figure}[!htb]
\centering
 \includegraphics[width=6cm]{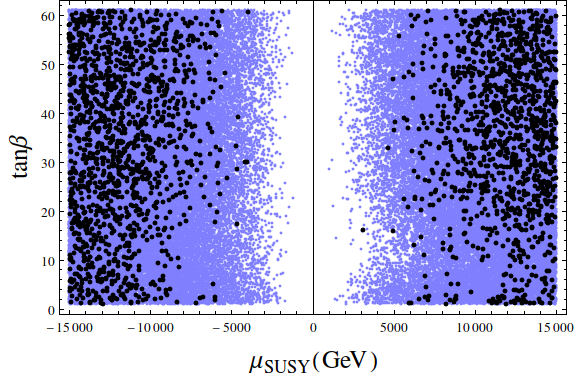}\includegraphics[width=6cm]{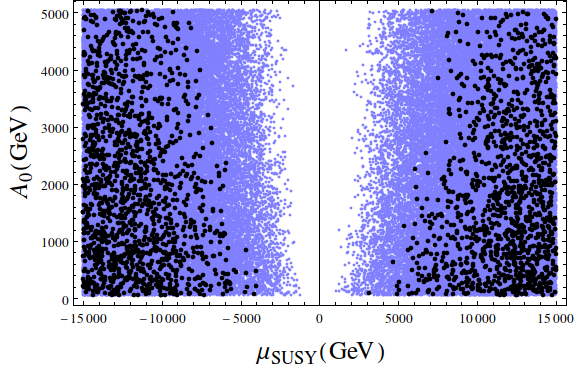}
 \caption{ Values of $\tan\beta$ (left) and $A_0$ (right) dependence on $\mu_{SUSY}$ for which the
   $a_{\mu}$ dis\-cre\-pan\-cy would get solved partially  by the LFV contributions (blue), or
   completely up to $1\sigma$ with the restriction $M_1<\frac{1}{3}\mu_{susy}$ (black). }
\label{graf-mu-vs-tanbmS}
\end{figure}
 
Figure ~\ref{graf-mu-vs-tanbmS} shows the relation of $\mu_{susy}$ with $\tan\beta$  and trilinear coupling $A_0$ for values for which the
  $a_{\mu}$ discrepancy receives contributions from the LFV
  terms. We see that there is a quite symmetrical behavior for any sign of $\mu_{susy}$. In order for the $a_{\mu}$ problem to be solved
  entirely by LFV $|\mu_{susy}|\gtrsim 4000~GeV$ and no restriction for $\tan\beta$.
  For smaller values of $A_0$ there will be less restriction on  $\mu_{susy}$.
 Although $\mu_{susy}$ values could be restricted by other sectors of the MSSM, {\em e.g.} the radiative
  corrections to the lightest Higgs mass
  \cite{Hahn:2013ria,Carena:2014nza}. On the other hand, there are
  other SUSY models, where the value of $\mu_{susy}$ could be
  naturally small \cite{Abe:2015xva}. \\

\begin{figure}
\centering
\includegraphics[width=6cm]{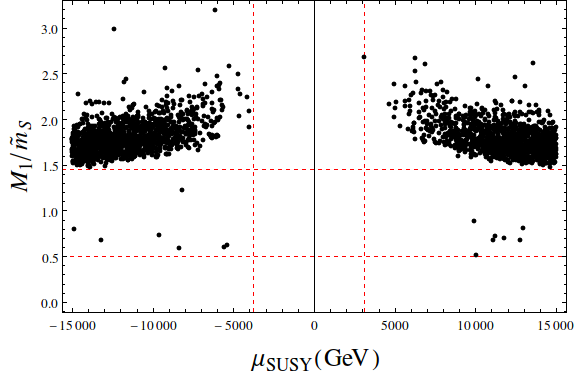}\includegraphics[width=6cm]{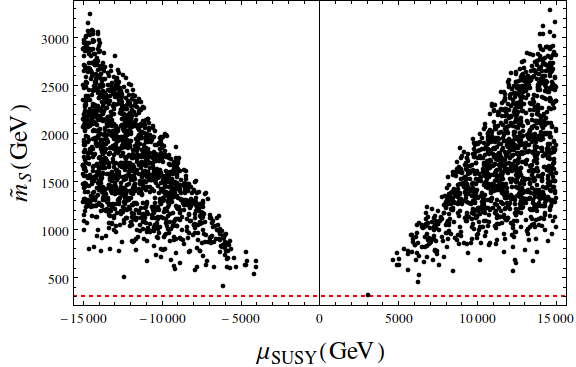}
\caption{The values for which the LFV contribution would explain completely the $a_{\mu}$
discrepancy  within theory and experimental data up to $1\sigma$, considering $M_1<\frac{1}{3}\mu_{susy}$. 
We show ratio on susy mass parameters $M_1/\tilde{m}_S$ (left) and $\tilde{m}_S$ (right), both with respect on $\mu_{susy}$ values.}
\label{grafg-2mSM1tB}
\end{figure}

Figure ~\ref{grafg-2mSM1tB}
  shows the ratio of the Bino mass $M_1$ with SUSY scalar mass $\tilde{m}_S$ where the points showed are solutions to $a_{\mu}$ discrepancy achieved up to $1\sigma$
  by the LFV contribution. We see a highly restricted regions for $1.5<M_1/\tilde{m}_S<2.5$, although we also have few points within $0.4\lesssim M_1/\tilde{m}_S<0.85$, 
  but there are no points for $0.8\lesssim M_1/\tilde{m}_S<1.2$. 
  We also see the behavior of these points  the scalar mass is
  highly restricted to the range of values $m_S\simeq [500, 3400]$~\rm{GeV},
  reaching the top values for larger values of  $|\mu_{susy}|$

We consider that in the region of parameter space where the points that solve completely the $g-2$ anomaly lie, the Bino-sleptons loop contribution will 
dominate over the chargino-sneutrino contribution.
Under this consideration is possible that the allowed parameter space is different from the MSSM with no FV terms in the charged lepton sector, 
where the chargino-sneutrino contribution is the dominant one \cite{Iwamoto:2013kla}.

\begin{figure}[!htb]
\centering
 \includegraphics[width=6cm]{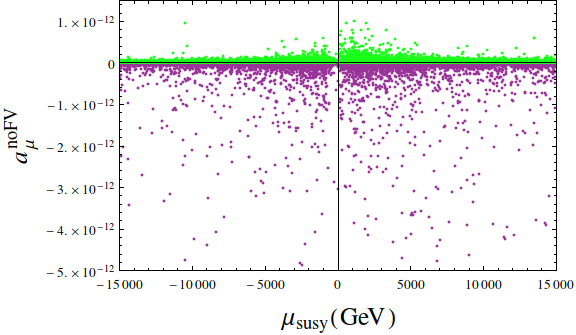}
 \caption{ Complete Bino-smuon loop contribution on MSSM  with no flavour violation to $g-2$, considering $A_0=0$ green points (lighter), and running $A_0$ for $(50, 5000)~GeV$ purple points (darker).}
\label{graf-g2NFV-vs-mS}
\end{figure}

\section{Summary and conclusions}

We proposed an ansatz for the trilinear scalar couplings considering a
two family flavour structure.  We obtain a non-universal slepton
spectrum and and slepton states are now flavor mixed. This specific family structure implies the possibility of lepton flavour
violation although avoids extra LFV contributions to $BR(\mu \to e \gamma)$.
In the method we used the FV is absorbed into the Lagrangian couplings instead of introducing a mass-insertion term into the propagator as used commonly 
in the literature. This method does not require a priori approximations  to reduce the loop amplitude integral expression.\\

We analyzed the parameter space
which gives values for these processes within experimental bounds.  
We considered that the lightest neutralino is mainly a Bino, specifically we consider the slepton-Bino loops. In order to have the Bino decoupled from Higgsino we take 
$M_1<|mu_{susy}|$. Under these
assumptions we showed that this FV couplings will include a mixture of four types
of sleptons running in the loop contributing to $a_{\mu}$, which in the interaction basis corresponds to
the smuons and the staus, as can be seen in Fig.~\ref{fig-slepton-cont}, and that for certain regions of parameter space it is
  possible to solve entirely the discrepancy between the experimental
  and theoretical values of $a_{\mu}$, in this case we specifically take a more restricted condition, $M_1<1/3|mu_{susy}|$. 
 The points that match with these conditions are given for the scalar SUSY mass scale $\tilde{m}_S$ involved in the LFV processes
  range between $450\lesssim\tilde{m}_S < 3300$ \rm{GeV}, the upper bound in the scalar mass is reached for
  $|\mu_{susy}|\sim 14~TeV$. 
  The possible Bino mass needed in order to solve the $a_{\mu}$ problem ranges
  from $\sim 350$ \rm{GeV} to $\sim 7.5$ \rm{TeV}, nevertheless the ratio of these masses is restricted to $0.4\lesssim \frac{M_1}{\tilde{m}_S}<3$,
  although we have very few point for  $\frac{M_1}{\tilde{m}_S}<0.9$, and the points around $\frac{M_1}{\tilde{m}_S}\sim 1$ are excluded.
  It is possible to contribute only partially to the
  $a_{\mu}$ problem, in which case a much larger parameter space is
  allowed (blue points). This partial contribution to $g-2$ will be important when the chargino-sneutrino contribution is included, 
  since it might change the allowed parameter space.
  This complete analysis we leave to a forthcoming work. Nevertheless, is worth mentioning again
   that it is natural to have differences in the parameter space as compared to the usual MSSM, where the chargino-sneutrino contribution is the dominant one.\\

It is interesting to notice that considering off-diagonal elements in the LR of the mass matrix block to be as large
as $~1$\rm{TeV} does not necessarily blow up the $BR(\tau \to \mu \gamma)$ process, 
instead, this assumption helps to reduce partially or completely  the $g-2$ discrepancy. 
In our case, we have considered off-diagonal terms in the soft trilinear couplings,
   of the order of $50~GeV < A_0 \lesssim $ 5\rm{TeV}.
We also compare our results with the no flavour violation $g-2$ MSSM one-loop contribution, where we obtain the same expressions given in the literature 
for complete calculation 
and in the numerical results we obtain small positive contributions to $g-2$ considering no contribution from the trilinear term $A=0$.    


\appendix
\section{Loop amplitude for $\tau\to \mu \gamma$}

We present here the expressions we obtain for the invariant amplitude
of the processes given in Fig.  1. For clarity in the expressions we have defined $g^2_{c}=\frac{ \rm tan^2\theta_{w}g^2_{1}}{16}$.
For general leptons in external particles represented by $i,j=e,\mu, \tau$, the diagram in Fig. 1 (a)
 we have
\begin{eqnarray}\label{eq:ama}
\mathcal{M}_{a}
&=&-eg^2_{c}\bar{u}(p_{1})\left[(S_{i}S_{j}- P_{i}P_{j})+(S_{i}P_{j}- S_{j}P_{i})\gamma^{5}\right]m_{\tilde{B}}\frac{1}{(2\pi)^{4}}\int dq^{4}\frac{2(p_{2}+q)\cdot \epsilon}{D_{q}D_{1}D_{2}}\nonumber\\
& &-eg^2_{c}\bar{u}(p_{1})\left[(S_{i}S_{j}+ P_{i}P_{j})+(S_{i}P_{j} + S_{j}P_{i})\gamma^{5}\right]\frac{1}{(2\pi)^{4}}\int dq^{4}\frac{2(p_{2}+q)\cdot \epsilon \sh{q}}{D_{q}D_{1}D_{2}}\ ,
\end{eqnarray}
where $D_{q}=q^{2}-m^{2}_{\tilde{B}}$, $D_{1}=(q+p_{1})^{2}-m^{2}_{\tilde{l}_{r}}$, $D_{2}=(q+p_{2})^{2}-m^{2}_{\tilde{l}_{r}}$, and $\epsilon$ is the photon polarization vector.
For the $\tau\to \mu \gamma$ decay, we have $i=\tau$ and $j=\mu$  and the  $S_{i,j}$, $P_{i,j}$ couplings are labeled as follows:   
$S_{i}= S_{\tilde{B} \tau \tilde{l}}$, $S_{j}= S_{\tilde{B} \mu \tilde{l}}$,  $P_{i}= P_{\tilde{B} \tau \tilde{l}}$ and  $P_{j}= S_{\tilde{B} \mu \tilde{l}}$. All the possible sleptons running inside the loop are indicated by the index
 $\tilde{l}= \tilde{\mu}_1,\tilde{\mu}_2,\tilde{\tau}_1,\tilde{\tau}_2$. The corresponding values are given in Table 1. For the anomaly $g-2$ we set $i=j=\mu$. \\
For the diagram Fig. 1(b)
we have
\begin{eqnarray}\label{eq:amb}
\mathcal{M}_{b} &=& -\bar{u}(p_{1})e\Sigma_{b} \frac{[\sh{p}_{1}+m_{i}]}{m^{2}_{j}-m^{2}_{i}} \sh{\epsilon} u(p_{2}),
\end{eqnarray}
with
\begin{eqnarray}
\Sigma_{b}
&=&m_{\tilde{B}}g^2_{c}\left[(S_{i}S_{j}- P_{i}P_{j})+(S_{i}P_{j}- S_{j}P_{i})\gamma^{5}\right]\frac{1}{(2\pi)^{4}}\int \frac{dq^{4}}{D_{q}D_{1}}\nonumber\\
& &+g^2_{c}\left[(S_{i}S_{j}+ P_{i}P_{j})+(S_{i}P_{j} + S_{j}P_{i})\gamma^{5}\right]\frac{1}{(2\pi)^{4}}\int \frac{dq^{4}\sh{q}}{D_{q}D_{1}}.\notag \\
\end{eqnarray}
The amplitude for Fig. 1(c) reads
\begin{eqnarray}\label{eq:amc}
\mathcal{M}_{c}
 &=&-\bar{u}(p_{1})e \gamma_{\mu}\epsilon^{\mu} \frac{[\sh{p}_{1}+\sh{k}+m_{j}]}{m^{2}_{i}-m^{2}_{j}} \Sigma_{c} u(p_{2}),
 \end{eqnarray}
 where
 \begin{eqnarray}
\Sigma_{c}
&=&m_{\tilde{B}}g^{2}_{c}\left[(S_{i}S_{j}- P_{i}P_{j})+(S_{i}P_{j}- S_{j}P_{i})\gamma^{5}\right]\frac{1}{(2\pi)^{4}}\int \frac{dq^{4}}{D_{q}D_{2}}\nonumber\\
& &+g^{2}_{c}\left[(S_{i}S_{j}+ P_{i}P_{j})+(S_{i}P_{j} + S_{j}P_{i})\gamma^{5}\right]\frac{1}{(2\pi)^{4}}\int \frac{dq^{4}\sh{q}}{D_{q}D_{2}}.\notag\\
\end{eqnarray}
The total amplitude which is the sum of Eqs.(\ref{eq:ama}, \ref{eq:amb}, \ref{eq:amc})  is written as follows:
\begin{eqnarray}\label{eq:mtotal}
\mathcal{M}_{T}&=&\bar{u}(p_{1})[\imath E_{ij}\sigma^{\mu\nu}k_{\nu}\epsilon_{\mu}+\imath F_{ij}\sigma^{\mu\nu}k_{\nu}\epsilon_{\mu}\gamma^{5}]u(p_{2})\nonumber\\
&=&\bar{u}(p_{1})[ \frac{E_{ij}}{2}+\frac{ F_{ij}}{2}\gamma^{5}][\sh{k},\sh{\epsilon}]u(p_{2})\ .
\end{eqnarray}
In the case of $i=\tau$ and $j=\mu$ we would have the expressions for $E_{ij}$ and $F_{ij}$ as in Eqs.(\ref{eq:etaumu}, \ref{eq:eftaumu}).\\

\section{The loop  contribution to the  muon anomaly}

The loop amplitude\footnote{ Notice that $g_{1}$ is the $U(1)$ coupling constant.} for the vertex correction is given by
\begin{align}\label{eq:primera}
\bar{u}(p_{1})\Gamma^{\mu}u(p_{2})&= \imath g_c \bar{u}(p_{1})\left[S_{\tilde{B}\mu,\tilde l}+P_{\tilde{B}\mu,\tilde l}\gamma^{5}\right]\frac{1}{(2\pi)^4}\int dk^{4}
\frac{\imath[\sh{k}+m_{\tilde{B}}]}{D_{t}}\imath \frac{\rm tan\theta_{w}g_{1}}{4}\nonumber\\
& \times [S_{\tilde{B}\mu,\tilde l}- P_{\tilde{B}\mu,\tilde l}\gamma^{5}]\frac{\imath}{D_{2}}\frac{\imath}{D_{1}}[2k+p_{1}+p_{2}]_{\mu}u(p_{2})\nonumber\\
&=\imath g^2_c \left[S_{\tilde{B}\mu,\tilde l}^2-P_{\tilde{B}\mu,\tilde l}^2\right]m_{\tilde{B}}\bar{u}(p_{1})\int \frac{dk^{4}}{(2\pi)^{4}}\frac{(2k+p_{1}+p_{2})^{\mu}}{D_{t}D_{1}D_{2}}u(p_{2})\nonumber\\
& +g^2_c \bar{u}(p_{1})\left[S_{\tilde{B}\mu,\tilde l}^2+P_{\tilde{B}\mu,\tilde l}^2\right]\int \frac{dk^{4}}{(2\pi)^{4}}\frac{(2k+p_{1}+p_{2})^{\mu}\sh{k}}{D_{t}D_{1}D_{2}}u(p_{2}) +\ldots
\nonumber\\&= g^2_c\bar{u}(p_{1})\left[(S_{\tilde{B}\mu,\tilde l}^2-P_{\tilde{B}\mu,\tilde l}^2)m_{\tilde{B}}B_{1}^{\mu}(q^2)\right]u(p_{2})\nonumber\\
& +g^2_c \bar{u}(p_{1})\left[ (S_{\tilde{B}\mu,\tilde
    l}^2+P_{\tilde{B}\mu,\tilde l}^2)B_{2}^{\mu}(q^2) \right]u(p_{2})
+ \ldots,
\end{align}
where $q^2=(p_{2}-p_{1})^2$ and the ellipsis means terms that are not  involved in the determination of the anomaly contribution. 
The propagators are given by
\begin{align}
Dt&=\frac{1}{k^2-m^{2}_{\tilde{B}}}\ ,\\
D_{1}&=\frac{1}{(p_{1}+k)^2-m^{2}_{\tilde{l}}}\ ,\\
D_{2}&=\frac{1}{(p_{2}+k)^2-m^{2}_{\tilde{l}}}\ .
\end{align}
By setting $q^2=0$ and considering that the muon mass is negligible compared to the supersymmetric particle masses inside the loop, 
the contributions to the anomaly  are found to be
\begin{align}
B_{1}^{\mu}(0)&=(p_{1}+p_{2})^{\mu}(b_{1}+b_{2})\ , \\
B_{2}^{\mu}(0)&=m_{\mu}(p_{1}+p_{2})^{\mu}(b_{2}+b_{4}),
\end{align}
where $m_{\mu}$ is the muon mass and  the scalar functions $b_{1,2,4}$ read as
\begin{align}
b_{1}&=-\frac{\imath}{(4\pi)^2} \frac{1}{\Delta_{lb}}\left[1+\frac{m^2_{\tilde{B}}}{\Delta_{lb}}\ln \left[\frac{m^2_{\tilde{B}}}{m^2_{\tilde{l}}}\right]\right]\ ,\\
b_{2}&=-\frac{\imath }{(4\pi)^2}\frac{1}{\Delta_{lb}}\left[-\frac{1}{2}+\frac{m^2_{\tilde{B}}}{\Delta_{lb}}-\frac{m^4_{\tilde{B}}}{\Delta_{lb}^2}\ln\left[\frac{m^2_{\tilde{l}}}{m^2_{\tilde{B}}}\right]    \right]\ ,\\
b_{4}&=-\frac{\imath}{(4\pi)^2}\frac{1}{\Delta_{lb}}\left[ -\frac{m^6_{\tilde{B}}}{\Delta_{lb}^3}\ln \left[\frac{m^2_{\tilde{l}}}{m^2_{\tilde{B}}}\right]-\frac{1}{2\Delta_{lb}}(m^2_{\tilde{B}}-\frac{2}{3}\Delta_{lb})
+\frac{m^4_{\tilde{B}}}{\Delta_{lb}^2}\right]\, ,
\end{align}
with $\Delta_{lb}=m^2_{\tilde{l}}-m^2_{\tilde{B}}$. Gathering all the pieces, the contribution of flavour violation to the muon anomaly reads
\begin{align}
a_{\mu}&
=\frac{ g^2_{c}m_{\mu}}{(4\pi)^2}\left[
(S^2_{\tilde{B}\mu,\tilde l}+P^2_{\tilde{B}\mu,\tilde l})\frac{m_{\mu}}{6m_{\tilde{l}}^2}F_1^N(x)
-(S^2_{\tilde{B}\mu,\tilde l}-P^2_{\tilde{B}\mu,\tilde l})
\frac{m_{\tilde{B}}}{3m_{\tilde{l}}^2}F_2^N(x)\right],
\end{align}
here $x=m_{\tilde{B}}^2/m_{\tilde{l}}^2$ and, for brevity we  define
$g^2_{c}=\frac{ \rm tan^2\theta_{w}g^2_{1}}{16}$.  We have used the notation for the functions $F_{1,2}^N(x)$ given in Ref.\cite{Stockinger:2006zn}.

\section*{Acknowledgements}
 
We are also grateful with the referee for very useful comments. We acknowledge very useful discussions with S. Heinemeyer. This work was partially supported by a Consejo Nacional de Ciencia y Tecnolog\'ia (Conacyt), Posdoctoral Fellowship
and SNI M\'exico. F.~F-B thanks the hospitality and support from Centro de Investigaci\'on en Ciencias F\'isico-Matem\'aticas,
Facultad de Ciencias F\'isico-Matem\'aticas, Universidad Aut\'onoma de Nuevo Le\'on. M.~G-B ac\-know\-led\-ges partial support from 
Universidad de las Am\'ericas Puebla. This work
was also partially supported by grants UNAM PAPIIT IN111115 and
Conacyt 132059.

\bibliographystyle{h-physrev5}
\bibliography{FlavioBibtex-MM}

\end{document}